# Droplets of the order parameter in a low density attracting electron system in the presence of a strong random potential

M.Yu. Kagan [1,4)], E.A. Mazur [2,3)ℵ]

[1)] National Research University Higher School of Economics, Myasnitskaya st., 20, 101000 Moscow, Russia

[2)] National Research Nuclear University MEPhI, Kashirskoye sh. 31, Moscow, 115409, Russia

[3)] Kurchatov Institute National Research Center, 123122, Moscow, ul. Academician Kurchatov, 1, Russia

[4)] P.L. Kapitza Institute for Physical Problems , Kosygin st., 2, 119334, Moscow, Russia

*E-mail: mkagan@hse.ru

ℵ eugen_mazur@mail.ru

The properties of a two-dimensional low density (n<<1) electron system with strong onsite Hubbard attraction U>W (W is the bandwidth) in the presence of a strong random potential V uniformly distributed in the range from -V to +V are considered. Electronic hoppings only at neighboring sites on the square lattice are taken into account, thus W=8t. The calculations were carried out for a lattice of 24x24 sites with periodic boundary conditions. In the framework of the Bogoliubov - de Gennes approach we observed an appearance of inhomogeneous states of spatially separated Fermi-Bose mixture of Cooper pairs and unpaired electrons with the formation of bosonic droplets of different size in the matrix of the unpaired normal states    We observed a decrease in the droplet size (from larger droplets to individual bielectronic pairs) when we decrease the electron density at fixed values of the Hubbard attraction and random potential. The obtained results are important for the construction of the gross phase diagram and understanding of the nature of the phase transition between superconducting, normal metallic and localized states in quasi-2D (thin) film of a dirty metal. In a more practical sense it is interesting also for the experimental implementation of superconducting qubits on quantum circuits with high impedances in granular superconductors.

## 1. Introduction

The influence of strong disorder on superconductivity has long been a subject of considerable interest both theoretically [1–7] and experimentally in thin films [19–20] and in granular superconductors used as a platform for superconducting qubits [21]. The generally accepted physical picture of the destruction of the superconducting state and the nature of the normal state has not been revealed yet. Part of the theoretical papers [1–7] suggests that the pairing amplitude Δ (r) is uniform in space (independent of r) even for a strongly disordered system. In [8–10], the Bogoliubov – de Gennes equations in real space were solved, however, quasibosonic limit of low electron densities was not fully investigated.

In present article, we consider strictly two-dimensional attractive-U Hubbard model for an s-wave superconductor at low temperatures T in a strong random potential and analyze this model in detail in the framework of the Bogoliubov-de-Gennes (BdG) approach, taking into account both the solutions with positive energy values $E_n$ and the solutions with negative values of the self-energy of the electronic system [11,12]. Our goal is to observe how the amplitude of local pairing Δ (r) changes spatially in the presence of disorder in the parameter ranges of strong Hubbard U>W and strong disorder V>W at low electron densities n=0.125-0.3 which have not yet been analyzed. We also pursue a goal to study the effect of spatial inhomogeneities on physically relevant correlation functions.

We note that recent experimental efforts to create superconducting qubits on quantum circuits with high impedances in granular superconductors [21] actualize again the problem of constructing gross phase diagrams and understanding the nature of the normal-superconductor and insulator-superconductor phase transitions in thin films of dirty metals [20,21] . In conclusion, we give some comments about the consequences of our results for the experiments on disordered thin films.

## 2. Equations for inhomogeneous case and formulation of the model

We consider the two-dimensional Hubbard model for an s-wave type disordered superconductor with short range (onsite) attraction between carriers, described by the Hamiltonian:

$$H = -t \sum_{i,j,\sigma} \left( c_{i,\sigma}^{+} c_{i,\sigma} + h.c. \right) + U \sum_{i} n_{i\uparrow} n_{i\downarrow} + \sum_{i,\sigma} (V_i - \mu) n_{i\sigma} . \qquad (1)$$

Here, the first term describes kinetic energy, where $t$ corresponds to hoppings to the neighboring sites of the square lattice, $U = -|U|$ is the amplitude of the Hubbard attractive potential at one site, $c_{i\sigma}^{+}$ ($c_{i\sigma}$) are the creation (annihilation) operators for electron with spin σ at the square lattice site $\mathbf{r_i}$, $n_{i,\sigma}$ - is the local electron density at site i for one component of the spin and μ is the chemical potential. The random potential $V_i$ is selected independently for each site and is governed by a uniform distribution in the range [−V, V]. Thus, V controls the intensity of diagonal disorder in the system.

We write the BdG equations for the Bogoliubov functions, that is, for the "bogolons" in the form of "electrons" $u_n(\mathbf{r_i})$ and in the form of "holes" $v_n(\mathbf{r_i})$ participating in pairing, [11,12]. Namely

$$-t \sum_{\pm\hat{\mathbf{x}},\pm\hat{\mathbf{y}},} u_n(\mathbf{r_i} \pm \hat{\mathbf{x}}, \pm \hat{\mathbf{y}}) + (U(\mathbf{r_i}) + V_i - \mu) u_n(\mathbf{r_i}) + \Delta_i \cdot v_n(\mathbf{r_i}) = E_n \cdot u_n(\mathbf{r_i}); \qquad (2)$$

$$\Delta_i^* \cdot u_n(\mathbf{r_i}) + t \sum_{\pm\hat{\mathbf{x}},\pm\hat{\mathbf{y}},} v_n(\mathbf{r_i} \pm \hat{\mathbf{x}}, \pm \hat{\mathbf{y}}) - (U(\mathbf{r_i}) + V_i - \mu) v_n(\mathbf{r_i}) = E_n \cdot v_n(\mathbf{r_i}). \qquad (3)$$

Here $E_n$ are the excitation energies in the system, $\mu$ is the chemical potential in the presence of a random diagonal disorder V. In order to take correctly into account both the pairing potential and chemical potential $\mu$ in the system, which requires taking into account states below the Fermi level, we will keep the solutions of system (1), (2) with both positive energy values $E_n$ and solutions with negative values of the self-energy of the electronic system. The most important contributions to the density of states of electrons $n(\mathbf{r_i}, E)$ at the lattice site $\mathbf{r_i}$ yields [11]:

$$n(\mathbf{r_i}, E) = 2 \sum_{n} \left[ |u_n(\mathbf{r_i})|^2 f_n + |v_n(\mathbf{r_i})|^2 (1 - f_n) \right] \delta(E - E_n), \qquad (4)$$

where $f_n = 1 / \left[ \exp(E_n / kT) + 1 \right]$ is the Fermi-Dirac distribution function for a given value of the excitation energy $E_n$. The chemical potential $\mu$ is determined from the self-consistency condition for the density of particles n (which corresponds to the average number of electrons per site ) :

$$\frac{1}{N} \sum_{i} n_i = 2 \sum_{i} \sum_{n} \left[ |u_n(\mathbf{r_i})|^2 f_n + |v_n(\mathbf{r_i})|^2 (1 - f_n) \right] = n. \qquad (5)$$

To obtain the normalization conditions for the eigenfunctions, we start from the relation [12]

$$\sum_n \left[ u_n(\mathbf{r}) u_n^*(\mathbf{r}') + v_n(\mathbf{r}) v_n^*(\mathbf{r}') \right] = \delta(\mathbf{r} - \mathbf{r}').$$

We assume that both $\mathbf{r}$ and $\mathbf{r}'$ lie within the cell of the radius $R_i$. We integrate over the cell $R_i$, assuming that within the cell $u_n(\mathbf{r}) = u_n(R_i)$, $v_n(\mathbf{r}) = v_n(R_i)$. Let $S_i$ be an area of the elementary cell, $S_i = \dfrac{S}{N}$. As a result, we obtain: $\sum_n \left[ u_n(R_i) u_n^*(\mathbf{r}') + v_n(R_i) v_n^*(\mathbf{r}') \right] S_i = \int_{S_i} dr \delta(\mathbf{r} - \mathbf{r}') = 1.$

It has been taken into account that in the process of integration over the cell $R_i$ by $\mathbf{r}$ at some point in the integration the variable $\mathbf{r}$ coincides with $\mathbf{r}'$. Now we integrate the resulting expression over $\mathbf{r}'$, assuming that within the cell $\sum_n \left[ u_n(R_i) u_n^*(R_i)^* + v_n(R_i) v_n^*(R_i) \right] S_i S_i = S_i$ so that

$\sum_n \left[ u_n(R_i) u_n^*(R_i)^* + v_n(R_i) v_n^*(R_i) \right] = = \dfrac{1}{S_i} = \dfrac{N}{S} = n_{knots}$, where $n_{knots}$ is the concentration of cells (nodes or sites) in the simplest case considered here. When we perform the space averaging within the unit cell of the type $\sum_n \left[ |v_n(\mathbf{r_i})|^2 \right]$, a multiplier $n_{knots}^{-1}$ appears which cancel the normalization on $n_{knots}$ in the right hand side of the normalization condition. Therefore, we can effectively write down the normalization condition in the standard form $\sum_n |u_n(\mathbf{r_i})|^2 + \sum_n |v_n(\mathbf{r_i})|^2 = 1.$ In the summation, both states with positive energy and with negative energy should be taken into account.

In (1), (2), in addition to the random potential $V_\mathbf{i}$, the mean field potential $U(\mathbf{r_i})$ at the site $\mathbf{r_i}$ is included [11,12]. This yields [30]:

$$U_i \equiv U(\mathbf{r_i}) = -U \sum_n \left[ |u_n(\mathbf{r_i})|^2 f_n + |v_n(\mathbf{r_i})|^2 (1 - f_n) \right]. \tag{6}$$

Introducing the Hartree-Fock shift of the chemical potential $\mu_i = \mu + U \dfrac{n_{el}(\mathbf{r_i})}{2}$, we rewrite (2,3) via the effective chemical potential $\mu_i$, which depends on site i:

$$-t \sum_{\pm \hat{\mathbf{x}}, \pm \hat{\mathbf{y}}.} u_n(\mathbf{r_i} \pm \hat{\mathbf{x}}, \pm \hat{\mathbf{y}}) + (V_i - \mu_i) u_n(\mathbf{r_i}) + \Delta_i \cdot v_n(\mathbf{r_i}) = E_n \cdot u_n(\mathbf{r_i}); \tag{7}$$

$$\Delta_i^* \cdot u_n(\mathbf{r_i}) + t \sum_{\pm \hat{\mathbf{x}}, \pm \hat{\mathbf{y}}.} v_n(\mathbf{r_i} \pm \hat{\mathbf{x}}, \pm \hat{\mathbf{y}}) - (V_i - \mu_i) v_n(\mathbf{r_i}) = E_n \cdot v_n(\mathbf{r_i}). \tag{8}$$

In (7), (18) $\mu_i$ has the form:

$$\mu_i = \mu + U \sum_n \left[ |u_n(\mathbf{r_i})|^2 f_n + |v_n(\mathbf{r_i})|^2 (1 - f_n) \right]. \tag{9}$$

In the same time the pairing potential $\Delta(\mathbf{r_i})$ appearing in (2), (3), (7), (8) reads [11,12]:

$$\Delta(\mathbf{r_i}) = U \sum_n u_n(\mathbf{r_i}) v_n^*(\mathbf{r_i}) (1 - 2 f_n). \tag{10}$$

Starting from the initial assumption for $\Delta_i(\mathbf{r_i}) \equiv \Delta_i$ and $\mu$, we numerically solved the BdG equations (2–10) for the eigenvalues $E_n$ and the eigenvectors $(u_n(\mathbf{r_i}))$, $(v_n(\mathbf{r_i}))$ on a finite 2D lattice of N × N sites with the periodic boundary conditions. Then we calculated the magnitudes of the amplitudes $(u_n(\mathbf{r_i}))$, $(v_n(\mathbf{r_i}))$ at the specific sites, and found the numerical values of the concentration of bogolons of the type u in the sum over the states $n_{u,\sigma}(\mathbf{r_i}) = \sum_n |u_n(\mathbf{r_i})|^2 f_n$ and of the type v : $n_{v,\sigma}(\mathbf{r_i}) = \sum_n |v_n(\mathbf{r_i})|^2 (1 - f_n) \neq n_{u,\sigma}(\mathbf{r_i})$ at low temperatures. Next, we calculated the local pairing amplitude $\Delta(\mathbf{r_i}) = U \sum_n u_n(\mathbf{r_i}) v_n^*(\mathbf{r_i})(1 - 2 f_n)$ at each site and the density values of the types u and v bogolons at the sites $\mathbf{r_i}$. We repeated the iteration process until the self-consistency for $\Delta_i$ and $n_i$ at each site was achieved.

### 3. The method of the calculations

In this work, we studied a lattice with 24 × 24 sites with the periodic boundary conditions. The calculation starts with random values $\Delta_i$ and $\tilde{\mu}_i$ on each of the sites, after which obtained Hamiltonian is numerically diagonalized using the library of programming languages maple and math-lab. As a result the eigenvectors $u_n(\mathbf{r_i}), v_n(\mathbf{r_i})$ and eigenvalues $E_n$ were found. Then $\Delta_i$ and $\tilde{\mu}_i$ are recalculated from the relations (10) and (9) and again were used as an input of a self-consistent cycle. The cycle is performed until the convergence of the values $\Delta_i$ and $\tilde{\mu}_i$ on each site with a given accuracy was achieved. The duration of the procedure on the NRC KI computer was several hours. Model (1) was studied for the parameter range $1 \le |U| / t \le 10$ and $0 \le V / t \le 12$ on a lattice of size N = 24 × 24 nodes. The study also considers low concentrations n = 0.125 at low temperatures. The value of the energy gap $E_{gap}$ is defined as the smallest positive eigenvalue of the energy. Correspondingly $\Delta_i$ is the value of the order parameter at the i-th site, V is the disorder amplitude. The values of $E_{gap}$ and $\Delta_{op}$, as well as all the results, unless otherwise specified, were averaged over 25 launches with various random mess distributions. The calculations also used the QUANTUM ESPRESSO software package version 6.2.1 [13,14], as well as the calculation visualizers.

### 4. Trends in the system with moderate disorder when we increase U

The first version of the calculations in the framework of BdG equations (2-10) was performed at a temperature close to zero on the grid N = 24x24 sites, at extremely low electron concentration n = 0.15, and in the limit of very strong attraction between the electrons U = -10t. The calculations were performed for moderate disorder V = 2.

Figure 1 shows the trends manifested with an increase in $|U|$ arising in the dependences of the spatial distribution of the electron density (left column), of the electron gap $\Delta$ (right column), and the value characterizing the coexistence of holes and electrons in real space (particle-hole mixing), middle column). We can see that the last quantity correlates with $\Delta$. The distribution of all the quantities with increasing U (localized at individual sites) is shown below:

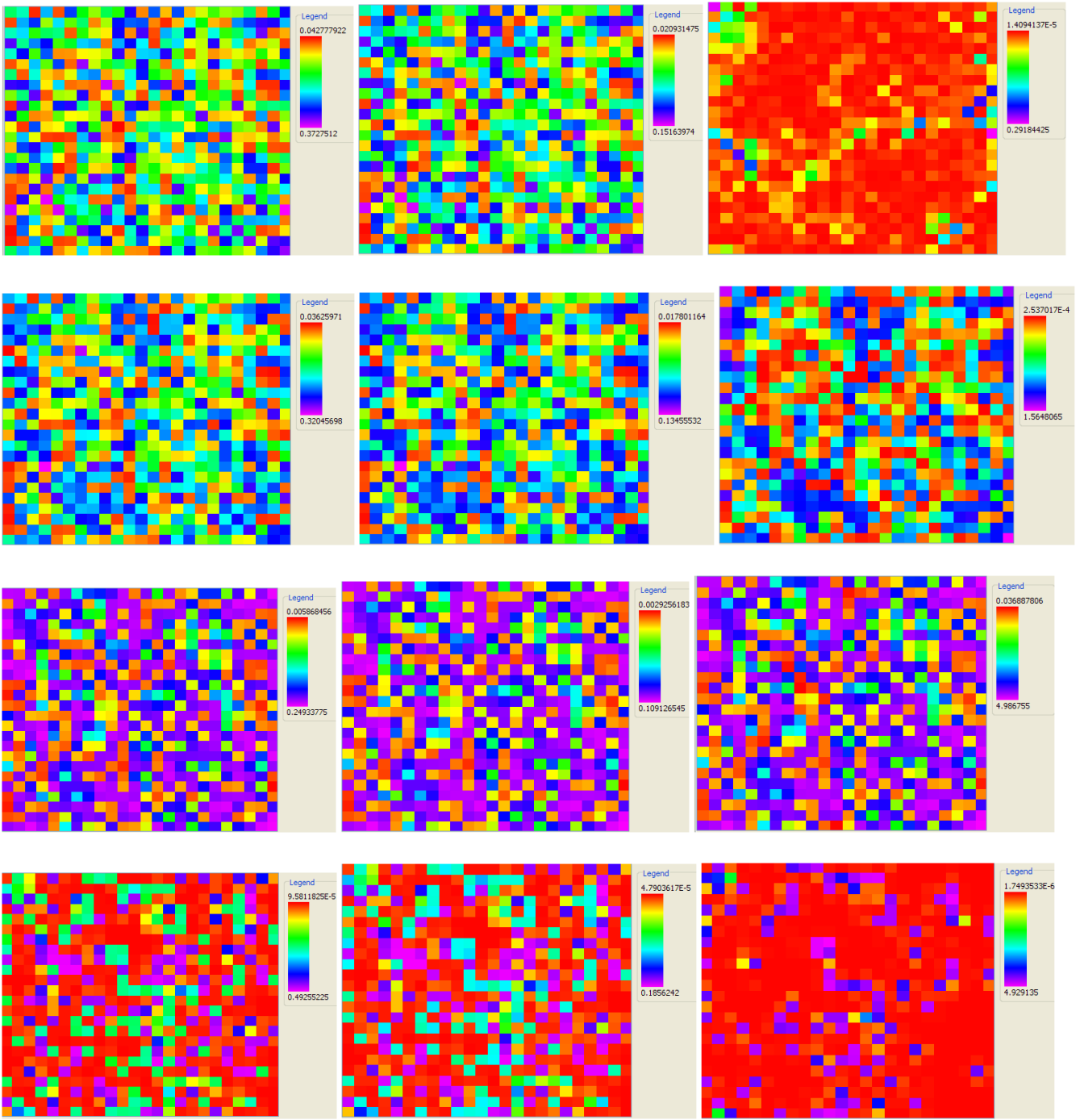

Fig. 1. Spatial distribution of the electron gap Δ (right column), electron density (left column) and the value characterizing the coexistence of holes and electrons in real space, middle column). The top line corresponds to $|U| = 2t$, V = 2t, the second line above corresponds to $|U| = 4t$, V = 2t, the third line above corresponds to $|U| = 10t$, V = 2t, the lowest line, presented here for comparison, corresponds to $|U| = 10t$, V=10t. The calculations were performed for low electron density n = 0.15.

As we can see from Fig. 1, at low electron concentrations n = 0.15, the system even at moderate values of disorder tends to form droplets of the order parameter, the spatial arrangement of which correlates with the spatial distribution of electron density in the system (right column), as well as with a value characterizing the coexistence of holes and electrons in real space. The electrons in the system are located in the regions with small values of disorder. The system is not an insulator under the considered parameters, since the distribution of electrons in space has not split into separate isolated regions. At low concentrations, the system exhibits the properties of a bosonic metal, the carriers of which are bosonic clusters consisting of several electron pairs, each of which is a boson.

To characterize the relative localization of pairs in Fig. 2 we show the graphs of the dependence of the correlator $\left\langle c_i^+ c_i^+ c_j c_j \right\rangle$, which is responsible for the transfer of pairs, on the distance as a function of concentration for given U and V. In the above calculations, we did not average the disorder configurations, all the results are presented for one disorder implementation.

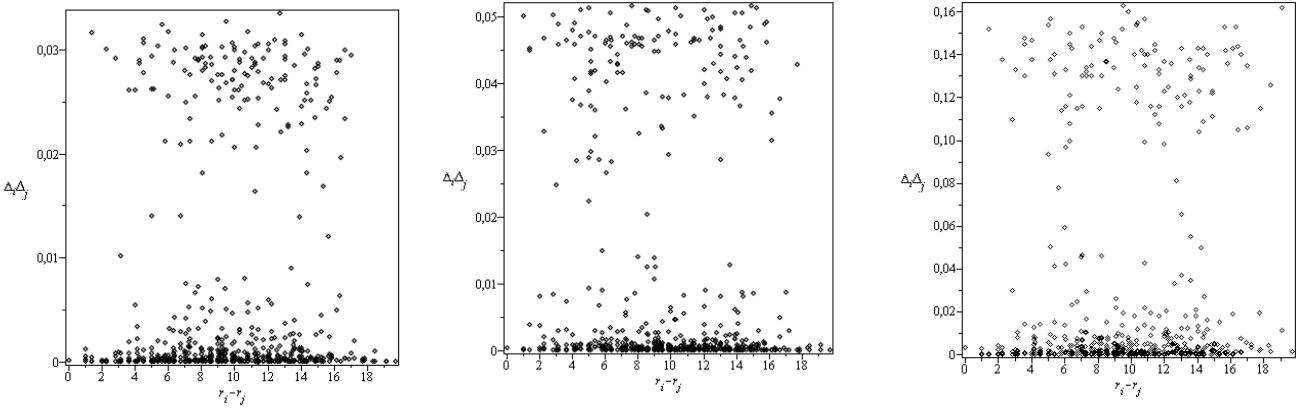

Fig. 2. The dependence of the correlator $\left\langle c_i^+ c_i^+ c_j c_j \right\rangle$, which reflects the spatial distribution of electron pairs, on the distance between the pairs, expressed in terms of the lattice constants. A series of graphs is presented with a variation in the average concentration for V = 8 and U = 6 a) n = 07; b) n = 0.30; c) n = 0.40.

As we can see from Fig. 2 c, the concentration value n = 0.31 corresponds to the proximity of the system to the transition to the state without pairing of electrons at all (with the pair parameter disappearing simultaneously at all sites). From Fig. 2 it follows that at all values of electron's concentration, a long-range order is established in the system, so that pairs located at a maximum distance from one another prevail (Fig. 2 a-c). As we can see from Fig. 2, the same spatial configuration of the mutual arrangement of the electronic pairs corresponds to the many values of the correlator quantity $\left\langle c_i^+ c_i^+ c_j c_j \right\rangle$. Thus, the system does not demonstrate a clear dependence of the correlator quantity $\left\langle c_i^+ c_i^+ c_j c_j \right\rangle$ on the mutual spatial arrangement of electron's pairs. An analysis of the behavior of the pair correlator (Fig. 2), which is responsible for the spatial distribution of pairs in the system, does not allow us to conclude that at these parameters the system is an insulator. We believe that in this section we have indicated certain limiting values of the parameters corresponding to the transition of the system to the state of the insulator.

## 5. Calculations for extremely low densities and weak disorder

Calculations similar to [8–10], however, for the previously unexplored region of extremely low electron concentration n = 0.125, were performed at a nonzero temperature on the grid N = 24x24 sites for U = -10t. The results of the previous section are supplemented by calculations for weak disorder with amplitudes V = 0.25; 2: 4.0; 10.0.

For U = 10 and V = 2, a series of figures (Fig. 3) shows how an increase in the concentration n affects the character and scale of the decrease in the average pairing parameter $\Delta_{op}$ and the gap value $E_{gap}$ with a concentration variation from n = 0.18 to n = 0.20. Results are averaged over five configurations.

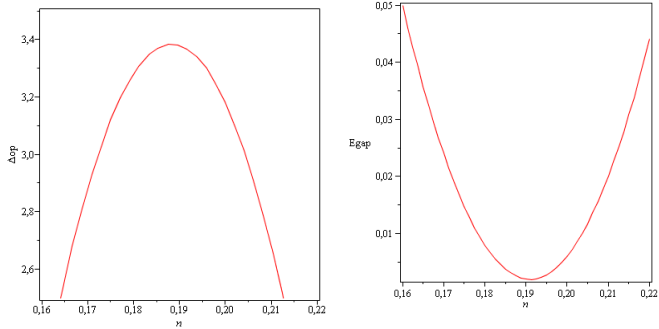

Fig. 3. The dependence of the average pairing parameter $\Delta_{op}$ and the value of the $E_{gap}$, on the electron density in the range from n = 0.18 to n = 0.20, U / t = -10, V = 2, temperature T = 0.01 for the lattice with 24x24 sites. The value of the order parameter $\Delta_{op}$ is shown (Fig. 3 a), as well as the value of the $E_{gap}$ (Fig. 3 b).

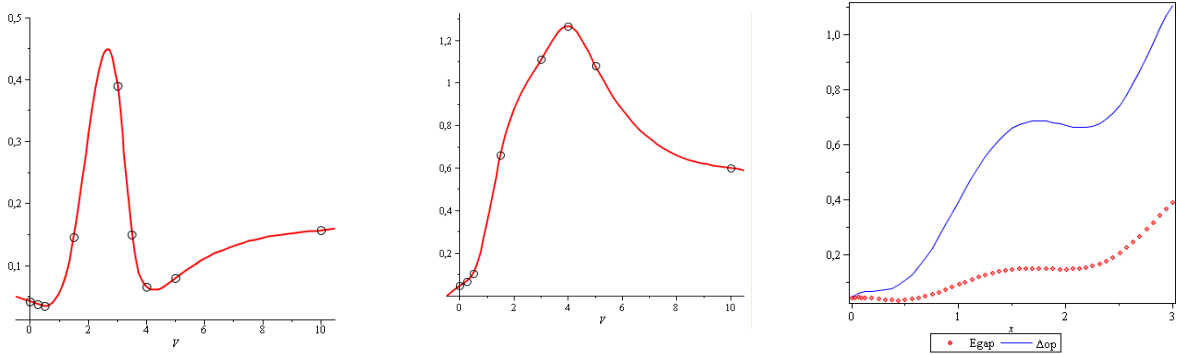

Fig. 4. a. Dependence of $E_{gap}$ on the amplitude of the disorder V; b. Dependence of $\Delta_{op}$ on the amplitude of the disorder V. n = 0.05, U / t = -8 for the lattice with 24x24 sites. In fig. 4. c. both graphs are shown more detailly within the same figure.

The presence of a minimum in the dependence of the $E_{gap}$ on the disorder degree V (Fig. 4.a) is consistent with the results from [8–10]. The following figures (Fig. 5-6) show the dependences of the spatial distribution Δ (left column), electron density (right column) and the value characterizing the coexistence of holes and electrons in real space (particle-hole mixing, middle column) for various amplitudes of disorder degree V in the system. For each particular picture we use its own contrast scale from white to black, so some configurations look faded compared to others due to the specific implementation of disorder. We start with one specific implementation of random distribution of the disorder amplitudes.

## 6. Calculations for low electron concentration, in the range of disorder from low to high values

This section presents the results of calculations in a wide range of the degree of disorder with the amplitudes V = 0.5; 6.0; 10.0 for a 24x24 system with a density of n = 0.15.

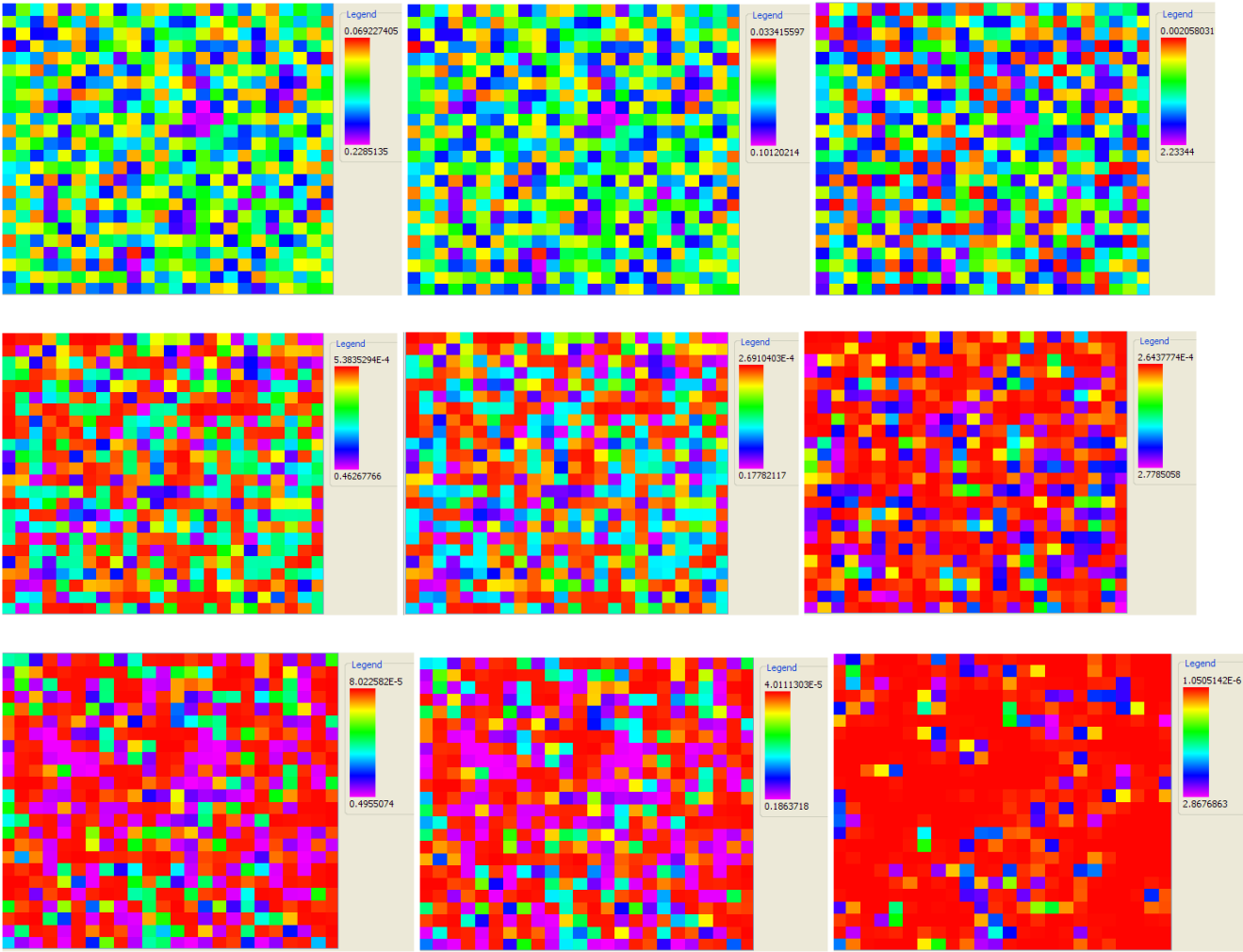

Fig. 5 Two-dimensional distribution of the order parameter (left column), electron-hole mixing (middle column) and electron density (right column) for n = 0.15, U / t = -6, on the 24 × 24 lattice with disorder amplitude a - upper line: V / t = 0.5; b - middle line V / t = 6.0; c - the bottom line V / t = 10.0.

Figure 5 shows that up to certain values of V (of the order of 5), correlations between the left and middle columns are clearly traced. Figure 5 shows the transition from the droplets of the order parameter to individual electron pairs - bosons with increasing disorder degree in the system. The question concerning the limiting value of the disorder at which our system passes from the Bose metal to the insulator in Fig. 5 remains open and requires additional investigation. Thus, our problem is divided into two independent problems: the Bose - metal problem and the insulator problem for the values of disorder exceeding the critical value, with dispersed droplets of the order parameter containing Bose states. From Fig. 5 c. from a comparison of all three columns, we can conclude that there is also a phase in which most of the electron pairs are localized, with a minority of delocalized pairs moving against the background of such an Bose insulating state.

Figure 6 shows the values of $\left| u_n\left(\mathbf{r}_i\right) \right|^2 + \left| v_n\left(\mathbf{r}_i\right) \right|^2$ with the positive energy values for the first three excitations for a series of values V = 0.5, 4.

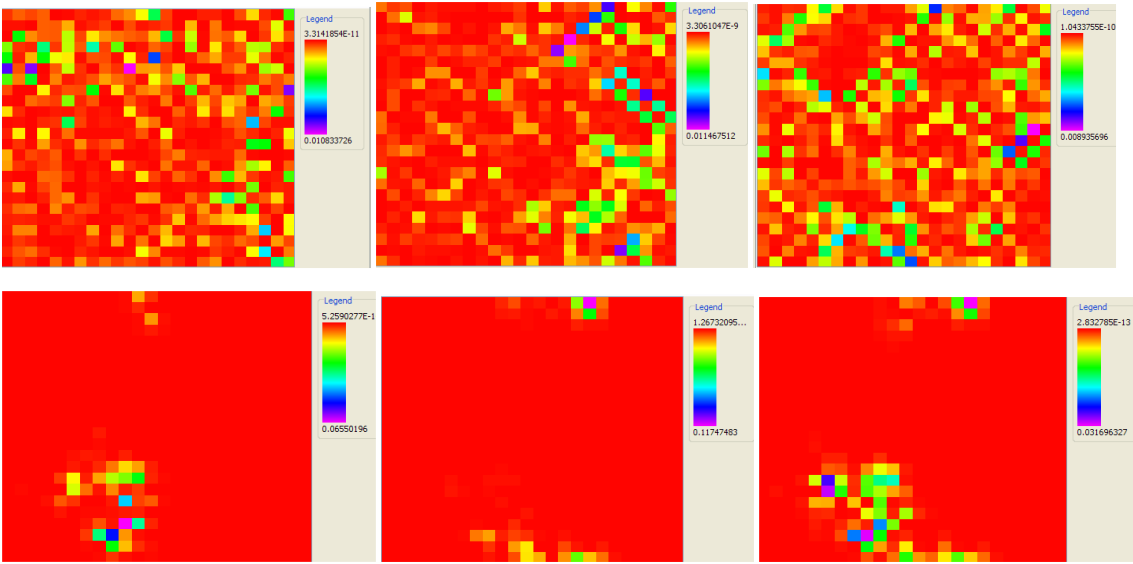

Fig. 6. The first three excitations are shown for the set of values V = 0.5; U = 6, T = 0.01, n = 0.3 (upper row), for the set of values V = 4; U = 6, T = 0.01, n = 0.15 (bottom row). From left to right, the excitation number for a given value of U increases, having energies of 0.000088, 0.004876, 0.013679 for the upper line and energies of 0.102606, 0.172969, 0.217828 for lower line, respectively.

## 7. Calculations for a high degree of disorder in the range of electron concentration from low n = 0.15 to average values n = 0.32

This section presents the results of the calculations in the range of electron densities from low n = 0.15 to moderate values n = 0.30 for the disorder amplitude of V = 8 and Hubbard amplitude U=6 for a lattice with 24x24 sites. Separate random excitations are also shown.

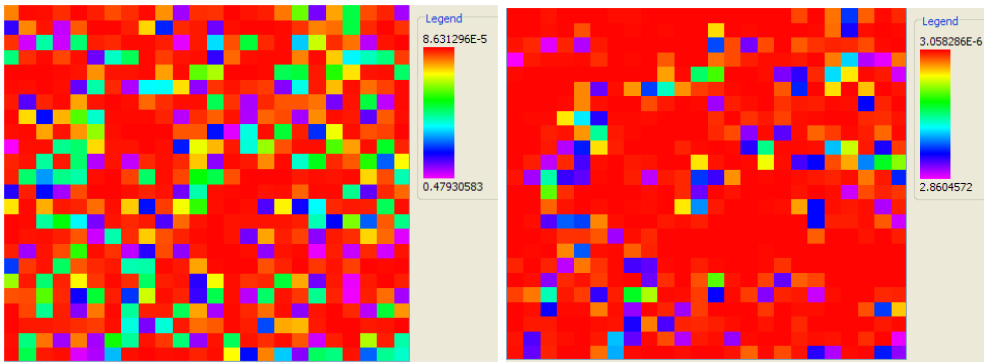

Fig. 7 Two-dimensional distribution of electron density (left column) and order parameter (right column) for n = 0.11, U / t = -6, on a 24 × 24 lattice with a disorder amplitude of V / t = 8.

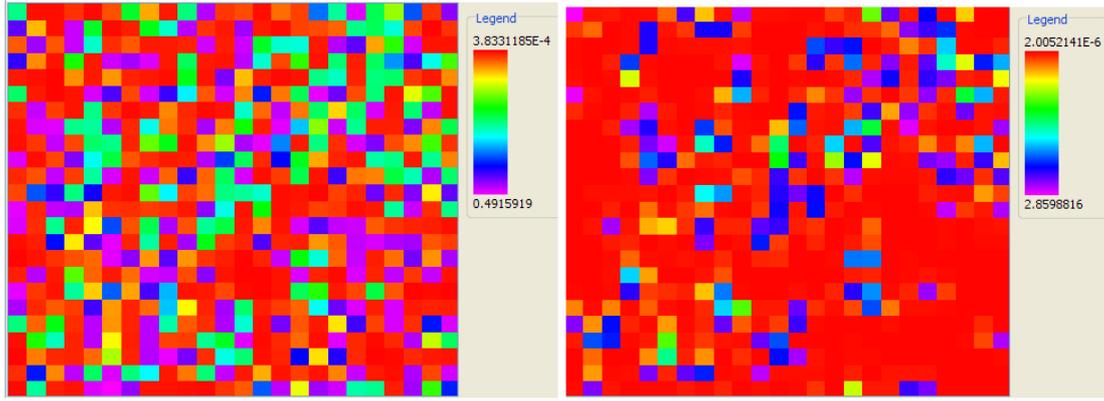

Fig. 8 Two-dimensional distribution of electron density (left column) and order parameter (right column) for n = 0.17, U / t = -6, on a 24 × 24 lattice with a disorder amplitude of V / t = 8.

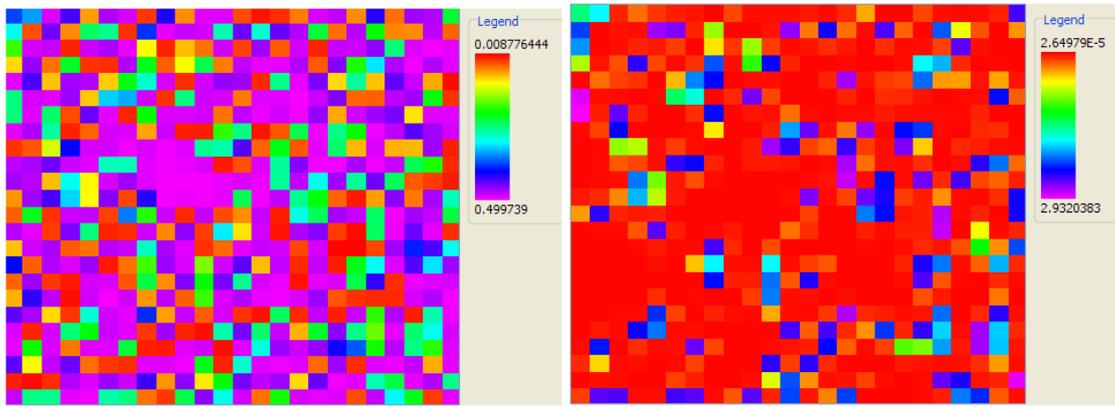

Fig. 9 Two-dimensional distribution of electron density (left column) and order parameter (right column) for n = 0.30, U / t = -6, on a 24 × 24 lattice with a disorder amplitude of V / t = 8.

As we can see, for the density n = 0.11, single pairs or doubled pairs of electrons (quartets) are observed on adjacent cells (Fig. 7), at n = 0.17, large droplets of the order parameter are observed (Fig. 8), and the electron distribution also has the form of droplets with a large number of electron pairs in each droplet. The peaks of the two-dimensional distribution of electron density and order parameter touch one another, but do not overlap. Droplets of the order parameter merge into the network of chains resembling the tree with a lot of branches (Fig. 8), which practically do not experience discontinuities along their length. At n = 0.30 the resulting network of the order parameter in this case is concentrated in the spatial regions — the "valleys" between the peaks of the electron density (Fig. 9).

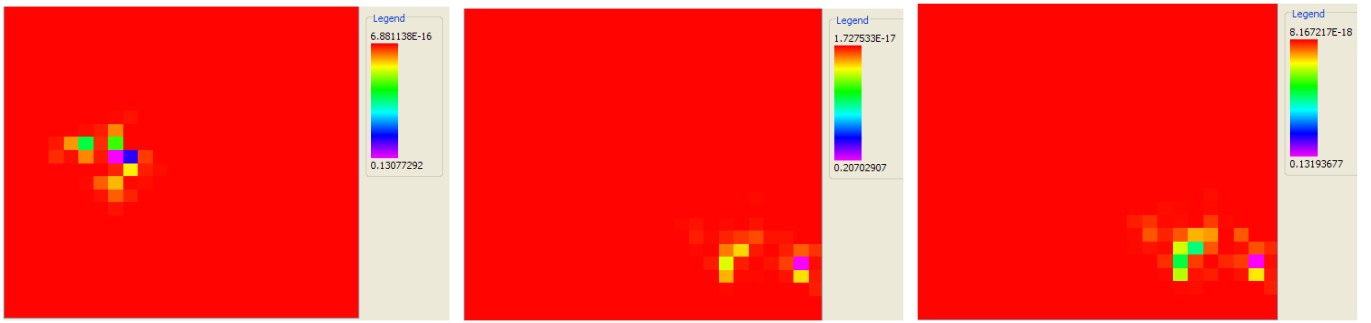

Fig. 10. The three lowest in energy excitations are shown for one particular realization of the disorder for the set of values V = 8, U = 6, n=0.28. From left to right, the excitation number increases for a given values of U, V, n:  a) excitation energy E1 = 0.070; b) excitation energy E2 = 0.0840, c) excitation energy E3 = 0.107.

## 8. Conclusions

As noted in earlier theoretical works [7–9, 24], the specific feature of the two-dimensional situation for the global phase diagram of Hubbard – Anderson type of models (with local onsite Hubbard interaction and diagonal disorder), as well as experimental studies of superconductivity and localization in dirty thin films [ 19,20], is connected with the fact that in quasi-2D systems besides the standard superconductor to normal metal phase transition in a certain range of parameters, a "direct" transition from a superconducting state to a dielectric localized state bypasssing the intermediate state of a normal metal is also possible. This circumstance makes the type of the calculations presented in this article highly relevant.

The main results of the work are as follows:

(1) As the disorder grows, the distribution P (Δ) of the local pairing amplitude Δ (r) becomes very wide, and even a significant weight develops near Δ ≈ 0, which can be interpreted as effective pairing in the system.

(2) The spectral gap in the single-particle density of states is maintained even with large disorder, despite the growing number of spatial nodes with Δ (r) ≈ 0. A detailed understanding of this effect arises from the study of the spatial variation of Δ (r) and eigenfunctions of the BdG equations. In our opinion, this effect needs to be revised by considering a model of the Fermi-Bose mixture type with the coexistence of pairs of electrons and single unpaired electrons.

We emphasize that our results were obtained in the strong coupling limit $|U|/t \gg 1$ and low density n << 1, and the limiting values of our calculations correspond to the cases of $|U|/t = 10$ and n = 0.125 for a two-dimensional square lattice with a bandwidth W = 8t. In this limiting case simple estimates using the formulas given in [1-3] give $\varepsilon_F \sim 0.8t, |E_b| \sim 0.2t$, respectively, for the Fermi energy and the binding energy of the electron pair. Thus $|E_b| < 2\varepsilon_F$, and we are still on the BCS side of the BCS - BEC crossover. If we introduce convenient notations for the binding energy, Fermi energy and band mass

$$|E_b| = \frac{1}{ma^2}, \varepsilon_F = \frac{k_F^2}{2m}, m = \frac{1}{2td^2}$$

, where d is the intersite distance , a is the radius of the pair, and $k_F$ is the Fermi momentum, then the dimensionless parameter for the BCS - BEC crossover related to the two-dimensional electron density $n_{2D} = \dfrac{k_F^2}{2\pi}$ for the parabolic spectrum takes the form

$$n_{2D}a^2 = \frac{k_F^2 a^2}{2\pi} = \frac{2\varepsilon_F}{|E_b|} \frac{1}{2\pi} \sim \frac{8}{6.28} \geq 1$$

, and for the limiting parameters of the problem it turns out to be close to 1.

This shows, as discussed in [1], that although we are on the BCS side of the crossover, the pairs are already quite compact, weakly overlap and almost touch each other. Thus, we are really very close to the limit when the pairs begin to "crush" each other, that is, to the Fermi-Bose limit of a mixture of compact pairs and unpaired single electrons.

Note that the model of the Fermi-Bose mixture is extremely rich, and can contain very diverse physics, in particular, with the possibility of the nanoscale phase separation in the system with the formation of spatially separated fermionic and bosonic clusters and many-particle droplets of different compositions (depending upon the sign and the relative magnitude of competing boson – boson, boson – fermion and fermion – fermion interactions $U_{BB}, U_{FF}, U_{FB}$, as well as the ratio of the densities of the two phases $n_B, n_F$ [22,23]).

We emphasize that all the results of this work were obtained in the Bogoliubov-de-Gennes approach, leading to a local minimum of the Free energy functional, without any additional approximations. This means, in particular, that a state with a more uniform spatial distribution of pairs in the system (a "uniformly mixed" Fermi-Bose solution of pairs and unpaired electrons) should correspond to a higher energy than the inhomogeneous state considered by us. A more rigorous consideration of the problem requires an analysis of the partial compressibilities of the Fermi-Bose mixture in order to prove rigorously the instability of the uniformly mixed state with respect to the nanoscale phase separation with the formation of droplets of one (bosonic paired) phase inside the matrix of another (unpaired) phase [23.25.27] . Such an analysis will be carried out separately together with the more rigorous determination of the nature of paired states inside the dropets (Bose metal or Bose insulator) as well as of the states in the unpaired matrix (normal fermionic metal or fermionic insulator).

Additional question in this context is connected with the phase-coherence and superconductivity. According to [21] one of the possible scenarios for superconducting transition in the spatially separated Fermi-Bose mixture  can be connected with the pair tunneling from one bosonic cluster to the neighboring one via the insulating fermionic barrier thus establishing macroscopic long-range order and phase-coherence in the system.

We also note that all the calculations in this paper were performed at a temperature close to zero, when the mean field theory works pretty well for a low density two-dimensional Fermi gas with attraction [1–4], and when the Berezinskii – Kosterlitz – Thouless fluctuation corrections [28–29] are small. The temperature evolution of the system in the quasi-bosonic region of parameters (the region of parameters corresponding to the appearance of a Fermi-Bose mixture) is also very interesting [1-3, 22, 25-26] and will be thoroughly investigated in subsequent works.

With an increase in the degree of disorder, a substantial decrease in the rigidity of superfluidity and diagonal correlations occurs, however, the amplitude fluctuations alone cannot destroy superconductivity. A simple analysis of this effective model within the framework of a self-consistent harmonic approximation leads to a transition to a nonsuperconducting state. It is expected that when "switched on" for the  droplets the mobility of such droplets will show the pseudogap properties in the spectrum.

In conclusion, let us discuss more detailly the evolution of the size of the droplets with the variation of the parameters of the problem. In particular, it is very instructive to fix the limiting ratio of strong Hubbard attraction $| U | / t = 10$ and to perform the gradual increase of the electron density, moving from $n = 0.125$ to larger densities close to half filling. In this case, our results show, that when we increase the density, at first the transition from single electron pairs to larger droplets (containing larger number of pairs) takes place. And than finally  a large percolating network of paired chains is formed in our system.

We also note that recent experimental efforts on various realizations of the superconducting flux qubits has raised an interest again in constructing a gross phase diagram and understanding the nature of the superconductor – normal and superconductor-insulator phase transition in the systems of reduced dimensionalities (D = 1.2). In this context, the experimental investigations of the quantum circuits with high impedance values in granular superconductors, and in particular in superconducting stripes of granular aluminum, which are alternative to ordinary Josephson media, seem to be very promising [24].

We are grateful to R.Sh. Ihsanov, E.A. Burovskii, K.I. Kugel, A.Ya. Tzalenchuk, A.S. Vasenko, N.N. Degtyarenko, A.V. Krasavin, A.A. Golubov for useful discussions of this work.

 This work was supported by the Competitiveness Enhancement Project of NRNU MEPhI (contract No. 02.a03.21.0005, 08.27.2013) using the equipment of the collective use center "Complex for modeling and processing data from research facilities of the mega-class" SIC "Kurchatov Institute" (subsidy of the Ministry of Education and Science, work identifier RFMEFI62117X0016), http://ckp.nrcki.ru/.

M.Yu.K. thanks for the support the HSE Program of Basic Research and expresses his gratitude to the RFBR fund (grant N20-02-00015).